\documentclass[useAMS,usenatbib]{mn2e}
\usepackage{graphicx}

\def\lesssim{\mathrel{\hbox{\rlap{\hbox{\lower4pt\hbox{$\sim$}}}\hbox{$<$}}}}
\def\gtrsim{\mathrel{\hbox{\rlap{\hbox{\lower4pt\hbox{$\sim$}}}\hbox{$>$}}}}

\def\LCDM{$\Lambda$CDM}
\def\hkpc{$h^{-1}{\ }{\rm kpc}$}
\def\hMsun{$h^{-1}{\ }{\rm M_{\odot}}$}
\def\nbody{$N$-body}
\def\IIM{\rm IIM}
\def\zform{$z_{\rm form}$}
\newcommand{\Table}[1]{Table~\ref{#1}}
\newcommand{\Sec}[1]{Section~\ref{#1}}
\newcommand{\Eq}[1]{equation~(\ref{#1})}
\newcommand{\Fig}[1]{Fig.~\ref{#1}}
\newcommand{\mlapm}{\texttt{MLAPM}}
\newcommand{\mhf}{\texttt{MHF}}

\def\ea{et~al.~}                            

\newcommand{\NewA}[3]   {\mbox{#3, NewA, #1, #2}}

\newcommand{\AJ}[3]     {\mbox{#3, AJ, #1, #2}}
\newcommand{\ApJ}[3]    {\mbox{#3, ApJ, #1, #2}}

\newcommand{\MNRAS}[3]  {\mbox{#3, MNRAS, #1, #2}}
\newcommand{\PASA}[3]   {\mbox{#3, PASA, #1, #2}}
\newcommand{\Nature}[3] {\mbox{#3, Nature, #1, #2}}

\title[Interactions \& Mass Loss from Satellite Galaxies in CDM Haloes]
      {The Importance of Interactions for Mass Loss from Satellite Galaxies in Cold Dark Matter Haloes}
      \author[Knebe~\ea]
       {Alexander Knebe$^{1}$, 
         Chris Power$^2$, Stuart P.~D. Gill$^{2,3}$, Brad K. Gibson$^{4, 5}$ \\
	  $^1$Astrophysikalisches Institut Potsdam, 
	     An der Sternwarte 16, 14482 Potsdam, Germany\\
        $^2$Centre for Astrophysics~\& Supercomputing, 
            Swinburne University, Mail \#39, P.O. Box 218, 
            Hawthorn, Victoria, 3122, Australia \\
        $^3$Columbia University, Department of Astronomy, 
		550 West 120th Street, New York, NY 10027, USA\\
        $^4$Laboratoire d'Astrophysique, Ecole Polytechnique Federale de Lausanne (EPFL), CH-1290, Sauverny, Switzerland\\
        $^5$Centre for Astrophysics, University of Central Lancashire, Preston, PR1 2HE, United Kingdom
       }

\date{Received ...; accepted ...}

\pagerange{\pageref{firstpage}--\pageref{lastpage}} \pubyear{2005}

\begin{document}
\label{firstpage}

\maketitle

\begin{abstract}

We investigate the importance of interactions between dark matter 
substructures for the mass loss they suffer whilst orbiting within a
sample of high resolution galaxy cluster mass Cold Dark Matter haloes
formed in cosmological N-body simulations. We have defined a quantitative 
measure that gauges the degree to which interactions are responsible for mass
loss from substructures. This measure indicates that interactions are more
prominent in younger systems when compared to older more relaxed
systems. We show that this is due to the increased number
of encounters a satellite experiences and a higher mass fraction in satellites.
This is in spite of the uniformity in the distributions of relative
distances and velocities of encounters between substructures within the 
different host systems in our sample.

Using a simple model to relate the net force felt by a single satellite 
to the mass loss it suffers, we show that interactions with other
satellites account for $\sim 30\%$ of the total mass loss experienced
over its lifetime. The relation between the age of the host and the importance
of interactions increases the scatter about this mean value from $\sim 25\%$
for the oldest to $\sim 45\%$ for the youngest system we have studied.
We conclude that satellite interactions play a vital role in the 
evolution of substructure in dark matter halos and that a significant
fraction of the tidally stripped material can be attributed to these
interactions. 

\end{abstract}


\setcounter{footnote}{1}

\section{Introduction}
\label{sec:intro}

It has been understood for some time that the structure of a galaxy can
be affected by tidal interactions with its close neighbour(s) 
\citep[e.g.][]{TT72}; tell-tale signs such as tidal tails and disturbed 
morphologies provide a visible record of these encounters. Around our own 
Galaxy, there is substantial evidence for its tidal interaction with the 
Small and Large Magellanic Clouds (SMC and LMC), the consequences of which 
have been studied in detail
\citep[e.g.][]{lin95,oh95,gardiner96,yoshizawa03,bekki05,mastropietro05,tim},
Furthermore, an increasing number of studies have uncovered evidence for 
tidal stripping -- in the form of stellar streams -- in the Galactic 
halo \citep[e.g.][]{Helmi,Ibata03}; these streams represent material that 
has been stripped from infalling satellites as they are disrupted by our 
Galaxy. The detection of such streams will become more commonplace in the 
coming years as the sensitivity of surveys improve 
\citep[e.g. ][]{Odenkirchen,Navarro}, but there are already examples of 
stellar streams further afield, such as around M31 \citep[][]{Ibata04}.
Moreover, \citet{Mihos2} recently reported the discovery of intracluster 
light in the Virgo cluster, revealing several long ($>$100 kpc) tidal 
streamers. 

These results represent compelling evidence that satellite galaxies tidally 
interact with their more massive hosts, and consequently lose some fraction of
 their mass. The effect of a satellite's interaction with its host and the 
mass loss it suffers has been studied in some detail \citep[e.g.][]{Hayashi}, 
and it can be argued that it is relatively well understood. In comparison,
the importance of a satellite galaxy's interactions with other satellite 
galaxies, the nature of these interactions and the contribution they make 
to its mass loss is less well understood. There is evidence to suggest 
that tidal interactions \emph{between} satellite galaxies occur; \citet{Zhao} 
and \citet{Ibata98} investigated whether the Sagittarius Dwarf galaxy could 
have experienced an encounter with the SMC and LMC some 2--3 Gyrs ago, while
the disturbed HI distribution noted by \citet{yun94} in the M81 group is 
highly suggestive of tidal interactions between the group galaxies. 
\citet{Goto} argues that tidal interaction between galaxies is the dominant 
mechanism driving cluster galaxy evolution and underpins the Butcher-Oemler 
effect and the morphology-density relation.\\ 

It has been understood for some time that dark matter haloes must play an 
important dynamical role in encounters between galaxies because they 
significantly reduce the merging timescale \citep[][]{B88}. Examples 
of tidal interaction and merging are observed in relatively low-density 
environments (i.e. the field), but how reasonable is it to expect that 
interactions should be more common in higher density environments such as 
galaxy groups and clusters? Tidal interactions have been proposed as a 
mechanism for galaxy transformation in galaxy clusters, such as the 
``harassment" scenario envisaged by \citet{MLK98}, but what does the favoured
paradigm for cosmological structure formation, the Cold Dark Matter model, 
predict? 

The aim of this paper is to quantify the importance of satellite-satellite 
encounters and to assess their impact upon the transformation and mass 
loss of the substructure population within the context of the Cold Dark Matter
(CDM) model. We have drawn on a sample of high resolution cosmological N-Body 
simulations of cluster mass dark matter haloes and analysed the interactions 
of the substructure haloes (hereafter subhaloes) both with the host halo and 
with other subhaloes. We associate these subhaloes with the hosts of satellite 
galaxies \citep[but see][]{gao04} and in what follows we use the terms subhalo
and satellite (galaxy) interchangeably. The fine time sampling of our 
simulations allow us to follow the time evolution of subhalo properties in 
detail, and so we can determine the relative contributions of the host and
 the other subhaloes to changes in a subhalo's structure.

In a previous study \citep[][]{KGG04} we quantified the frequency of
encounters between subhaloes orbiting within a common CDM cluster mass
halo, considering the period between the halo's formation
 redshift\footnote{We defined this
to be the redshift at which the mass of the most massive progenitor was
half the system's present day mass; this was typically $z \sim 0.5$ for
the haloes we examined.} and the present day. We found that, on
average, $30\%$ of the ``satellite galaxy" population experienced at
least one encounter per
orbit with another satellite galaxy. This result was sensitive to the age of
the host halo, with a clear trend for more encounters in younger
systems. We also reported a correlation between the number of 
encounters and halocentric radius -- satellite galaxies closer to the 
centre of the host were measured to experience more interactions,
although we note that this simply reflects the increasing spatial
density of satellites with decreasing radius within a host halo.

The principal shortcoming of the approach adopted in \citet{KGG04} is
that we neglected the relative velocities of the satellite galaxies;
our satellites may have experienced encounters, but we had no
information about their specific nature, i.e. were they fast
or slow? Such information is important when considering the impact on
the satellite's structure. In this present study, we elaborate on that work 
by including information about the relative velocities of
the satellites. In other words, we can now estimate the 
\emph{importance} of encounters in addition to the frequency with which
they occur, allowing us to differentiate between \emph{slow}
encounters, which we expect to be extremely disruptive to the satellite
structure, and \emph{fast} encounters, whose impact are likely to be minimal. 
We define a quantitative measure for interactions, which we 
call the \emph{integral interaction measure}, based 
upon the force acting on a satellite over a given period of time,
i.e. the (induced) momentum change. Whereas before we could examine the
number of encounters a satellite galaxy experienced per orbit, we may
now study how the instantaneous force due to encounters acting on a 
satellite galaxy varies along its orbit and how this correlates with
mass loss, thus providing a natural measure of the importance of mutual
interactions between satellite galaxies.

In what follows, we motivate our choice of the integral interaction
measure (hereafter IIM) as a gauge for the importance of interactions
between satellite galaxies, and present the results of our
analysis of a suite of high resolution cluster mass haloes that formed
assuming the \LCDM~cosmology. We demonstrate the suitability of the IIM
for our purposes by performing a series of experiments with ``cleaned''
simulations, in which we track the detailed mass loss history of a
single satellite galaxy in a host halo in which the substructure has 
been removed. Finally, we compare and contrast our results with those
of previous studies, and comment on their observable consequences.

\section{The Simulations}
\label{simulations}

Our analysis is based on a suite of eight high-resolution \nbody\
simulations \citep[][]{GKG04a, GKGD04} carried out using the
publicly available adaptive mesh refinement code \mlapm\ \citep[][]{KGB01}
 in a standard \LCDM\ cosmology ($\Omega_0 =
0.3,\Omega_\lambda = 0.7, \Omega_b h^2 = 0.04, h = 0.7, \sigma_8 =
0.9$). Each run focuses on the formation and evolution of a dark
matter galaxy cluster containing of order one million particles, with
mass resolution $1.6 \times 10^8$ \hMsun\ and force resolution
$\sim$2\hkpc\ which is of the order 0.05\% of the host's virial
radius. These simulations have the required resolution to follow the
satellites within the very central regions of the host potential
($\geq$5--10\% of the virial radius) and the time resolution to
resolve the satellite dynamics with good accuracy ($\Delta t
\approx$170~Myrs). Such temporal resolution provides of order 10-20
timesteps per orbit per satellite galaxy, thus allowing these
simulations to be used in a previous paper \citep[][]{GKGD04} to
accurately measure the orbital parameters of each individual satellite
galaxy.

Substructure within these haloes is identified using the halo finder
\mhf~(\mlapm's-halo-finder). \mhf~is based upon the \nbody\ code \mlapm\
and acts with exactly the same accuracy as the \nbody\ code itself; it
is therefore free of any bias and spurious mismatch between simulation 
data and halo finding precision arising from numerical effects. 
We applied \mhf\ to each of our eight host halos at their formation
time which is the redshift \zform\ where the halo contains half of its 
present day mass. We track the orbits of each of the satellites 
identified within and around the host halo from \zform until $z=0$ and
follow the evolution of their properties in great detail. For further
details relating to the properties of the satellite galaxies, we refer
the reader to the \citet{GKG04a,GKGD04,GKG05} series of papers.

\section{The Analysis}
\label{Analysis}

In what follows, we have considered only those satellites that have
completed at least one full orbit within their host halo, corresponding to
more than 70\% of the subhaloes. The number distribution of orbits peaks at 
about 1--2 orbits with the tail extending to as many as four orbits for the 
older host halos. A detailed discussion of the orbital properties (amongst 
others) of the substructure population can be found in \citet{GKG04a}.

We restricted our sample of satellites to those that contain at least 100 
particles, which translates to a minimum mass of $M_{\rm sat} \geq 2 
\times 10^{10}$ \hMsun. To ensure that our results are not affected by 
resolution effects, we checked that all results presented below are 
recovered when the lower mass limit is gradually increased; that is,
we considered additional lower mass cuts corresponding to 200 and 400
 particles and we can confirm that our results are unaffected.

\begin{figure}
\begin{center}
\includegraphics[width=8cm]{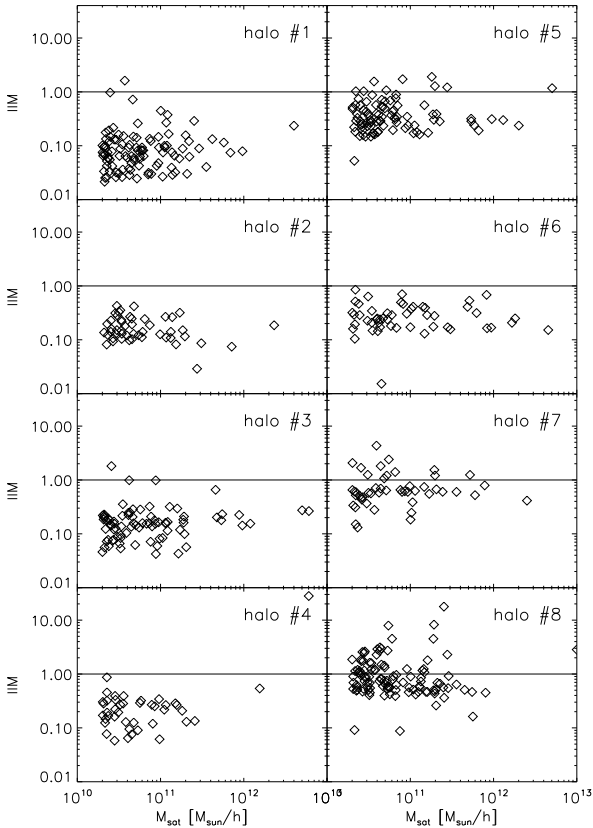}
\caption{The integral interaction measure as a function of satellite mass.}
\label{MsatIIM}
\end{center}
\end{figure}

\subsection{Integral Interaction Measure} \label{IIMmeasure}

We begin by calculating the forces acting on each satellite galaxy $i$
for each available snapshot of the simulation, treating it as a point 
particle with mass $m$. The force $F_{\rm host}^i$ exerted by the 
host halo and the force $F_{\rm sat}^i$ exerted by all other satellites
are given as follow

\begin{equation}
\begin{array}{rll}
F_{\rm sat}^i  & = & \displaystyle G m_i \sum_{i\ne j} \frac{m_j}{|r_i-r_j|^2}\\
\\
F_{\rm host}^i & = & \displaystyle G m_i \frac{M_{\rm host}(< r_i)}{r_i^2} \ , \\
\end{array}
\end{equation}

\noindent
where $m_i$ is the mass of satellite $i$ and $M_{\rm host}(< r_i)$ the mass
of the host interior to the satellite distance $r_i$. We need to stress that
both these formulae assume spherical symmetry and hence are only approximations
to the ``true'' forces.

\noindent
We define a so-called (dimensionless) ``integral-interaction-measure''
-- IIM -- for each individual satellite galaxy as follows:

\begin{equation} \label{IIMintegral}
 \IIM^i = \frac{1}{T} \int_{0}^{T} 
       \frac{F_{\rm sat}^i(t)}{F_{\rm host}^i(t)} dt
\end{equation}

\noindent
where we integrate over a time interval $[0,T]$, which is the time 
satellite $i$ has spent within its host's virial radius. Here we also note that
due to our definition the IIM values scale linearly with the ``average 
satellite mass". The discrete nature of the time sampling of our data requires
that the integral should be expressed as the following summation:

\begin{equation} \label{IIM}
 \IIM^i = \frac{1}{t_{\rm now} - t_i} \sum_{t=t_i}^{t{\rm now}} 
       \frac{ F_{\rm sat}^i(t_{m}) }{ F_{\rm host}^i(t_{m}) }\Delta t
\end{equation}

\noindent
where $t_{\rm now}$ is the age of the Universe at redshift $z=0$, 
$t_i$ the age of the Universe when the satellite enters the host halo,
and $\Delta t$ the time difference between two consecutive outputs. 
We average the forces exerted by both the other satellites and 
the host halo over the consecutive outputs, i.e. 
[$t-\Delta t/2, t+\Delta t/2]$, or ``mid-point integration'' of 
\Eq{IIMintegral}:

\begin{equation} \label{ForceMean}
\begin{array}{rll}
   \displaystyle F_{\rm sat}^i(t_{m})   & = & 
   \displaystyle   \frac{1}{2} \left( F_{\rm sat}^i(t-\frac{\Delta t}{2}) + F_{\rm sat}^i(t+\frac{\Delta t}{2}) \right)  \\
\\
   \displaystyle F_{\rm host}^i(t_{m}) & = & 
    \displaystyle  \frac{1}{2} \left( F_{\rm host}^i(t-\frac{\Delta t}{2}) + F_{\rm host}^i(t+\frac{\Delta t}{2}) \right) \\
\end{array}
\end{equation}

\noindent
The integral-interaction-measure \Eq{IIM} can now be used as a
quantitative measure for the relative strength of satellite-satellite
encounters.

\subsubsection{Application of the Integral Interaction Measure} \label{SecIIM}

In \Fig{MsatIIM} we present the integral-interaction-measure $\IIM$, as 
defined by \Eq{IIM}, for each satellite in our suite of eight host
halos plotted as a function of satellite mass. This figure
suggests that there is no clear trend for interactions to correlate
with mass, as we might have expected; it would be rather surprising to 
find that, for instance, high-mass satellites tend to interact more 
prominently than low-mass ones (or vice versa). 

The most striking feature of \Fig{MsatIIM} is the apparent rise
of the $\IIM$ values as a function of decreasing age for the host
halos: the halos are ordered in age with halo~\#1 being 8.3 Gyrs old
and halo~\#8 a mere 3.4 Gyrs. 

\begin{figure}
\begin{center}
\includegraphics[width=8cm]{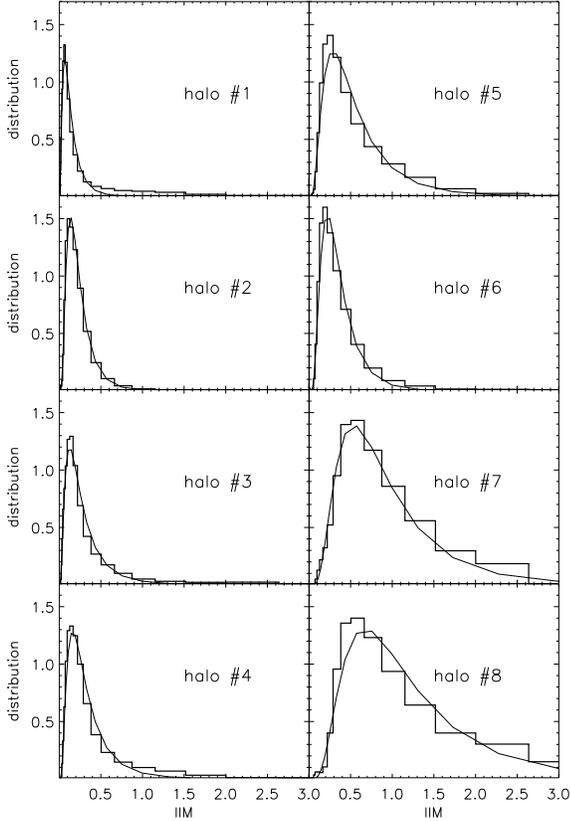}
\caption{Frequency distribution of the integral interaction measure.}
\label{NIIM}
\end{center}
\end{figure}

This can be better viewed in \Fig{NIIM} where we plot the distributions
of the integral interaction measure IIM. For all our eight host halos
these distributions have been fitted with a log-normal function

\begin{equation} \label{lognormal}
 n({\rm IIM}) = \displaystyle \frac{1}{x \sqrt{2\pi \sigma_0^{2}}} \exp
\left( {-\frac{\ln^2 ({\rm IIM}/{\rm IIM}_0)}{2 \sigma_0^2}} \right) ;
\end{equation}

\noindent
corresponding best fit parameters along with the halo age are listed in
\Table{AgeIIM} where IIM$_{\rm peak}$ = IIM$_{0} \exp(-\sigma^{2})$ for
a log-normal distribution.

\begin{table}
\caption{Best fit parameters of IIM distribution to lognormal.}
\begin{center}
\begin{tabular}{ccccc}\hline
halo         & age [Gyr] & IIM$_{\rm peak}$ & IIM$_{0}$ & $\sigma_0$ \\
 \hline \hline

\#       1 &       8.30 &       0.067 &       0.118  &      0.756 \\ 
\#       2 &       7.55 &       0.137 &       0.199 &       0.614 \\ 
\#       3 &       7.16 &       0.126 &       0.227 &       0.767 \\ 
\#       4 &       7.07 &       0.162 &       0.270 &       0.716 \\ 
\#       5 &       6.01 &       0.287 &       0.464 &       0.692 \\ 
\#       6 &       6.01 &       0.221 &       0.307 &       0.575 \\ 
\#       7 &       4.52 &       0.535 &       0.789 &       0.623 \\ 
\#       8 &       3.42 &       0.672 &       1.021 &       0.646 \\ 

\end{tabular}
\end{center}
\label{AgeIIM}
\end{table}%




The increase of the IIM with decreasing age of the host is consistent 
with the behaviour observed in \citet{KGG04}, in which it was noted that
the tail of the distribution of the number of encounters per orbit 
extended to larger values for younger host systems. However, the result
implied by the IIM is distinct from that presented in \citet{KGG04}
in the sense that we are considering the net force acting on a subhalo
over some time interval, whereas we previously considered encounters as
events in which a pair of subhaloes were spatially coincident. This
raises the question of whether or not the IIM is a reasonable measure of
interactions, and in particular, if it could simply be the case that it
is dominated by single encounters.

\begin{figure}
\begin{center}
\includegraphics[width=8cm]{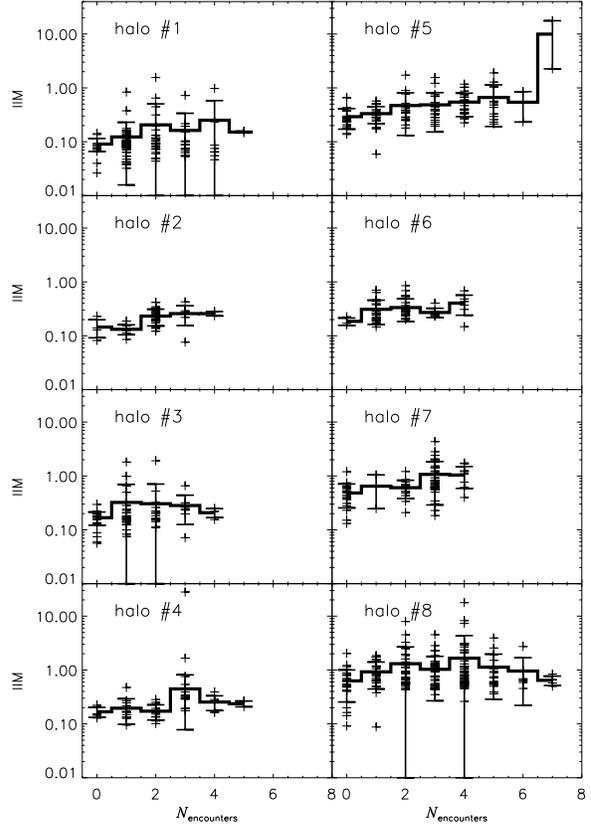}
\caption{Correlation of integral interaction measure IIM with the
         number of encounters as defined in the text.}
\label{encIIM}
\end{center}
\end{figure}

We investigate this in \Fig{encIIM}, where we examine the correlation 
between the IIM and the number $N_{\rm enc}$ of ``tidal encounters'' as
quantified by calculating the tidal radius of a given satellite 
\textit{induced by one of the other satellites} \citep{KGG04}.
Whenever the tidal radius becomes smaller than the
radius\footnote{We define the radius of a satellite either to be the 
virial radius, i.e. the radius where the mean averaged density (measured in
 terms of the cosmological background density $\rho_b$) drops below 
$\Delta_{\rm vir}(z)$, or the truncation radius, i.e. the point where the 
satellite's density profile
rises again due to the embedding in the host halo.} of the satellite we
 increment a counter $N_{\rm enc}$ for
that particular satellite that keeps track of the number of (perturbing) 
interactions with companion satellite galaxies.

\Fig{encIIM} clearly indicates that there is little (if any)
correlation between the number of satellite-satellite encounters
and the integral interaction measure \textit{for a single
satellite}. Moreover, also the scatter about the mean IIM value in each $N_{\rm enc}$ bin is not affected by the actual number of encounters experienced by the satellite.  This strongly suggests that the IIM value is not dominated 
by single events but rather is a cumulative quantity that is accrued
over the lifetime of a satellite. However, we stress that there is a 
correlation between the width of the distribution of encounters per
orbit \citep[c.f. Fig.3 in][]{KGG04} and the peak IIM value; although the
 IIM is not driven by single violent encounters, the greater the number of
 such events, the higher the IIM of the satellite. However, these arguments 
are based upon the assumption that the ``strength" of individual encounters 
is more or less equal. It still appears possible for \textit{one single 
strong} encounter to dominate the value of IIM. 

Another factor possibly affecting the observed correlation between
host halo age and interaction measure IIM is the mass fraction of satellites.
A simple check indicates that the younger the host the higher the fraction
of mass locked-up in satellite galaxies. This suggests that the IIM values
may in fact be influenced by the most massive subhaloes. We will come back to
 this point later in \Sec{TestCase} but can already confirm that the 
distributions presented in \Fig{NIIM} practically remain unaltered if we 
discard all satellites less massive than 1\% of the host's virial 
mass\footnote{One needs to bear in mind that the mass
spectrum of subhaloes extends down to as low as $10^{-4}$ -- $ 10^{-5} 
M_{\rm vir, host}$ \cite[e.g.][]{DeLucia}}, denoting the importance of massive
 subsystems.

In addition, we have investigated whether or not there exists a
relation between then IIM and either the eccentricity of a satellite's 
orbit or its pericentric distance, but we do not find strong evidence
for such a correlation. Although we observed a significant drop in the
number of encounters per orbit with increasing distance from the host's
centre, we find no comparable result for IIM. This indicates that
satellites ``encounter" each other with greater frequency closer to
the centre of the host, but that such encounters occur with high
relative velocities and so cause little structural damage. \emph{Encounters
in the central regions are therefore no more damaging than those in the
outer regions.} We elaborate upon this in greater detail in the following
 section.

\begin{figure}
\begin{center}
\includegraphics[width=8cm]{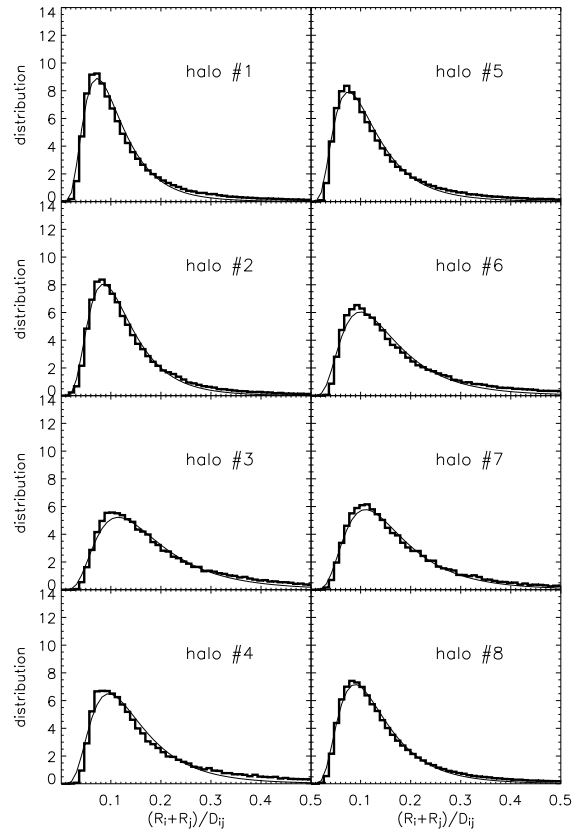}
\caption{Distribution of the distance of two satellites normalized by the
 sum of their individual radii; data from each host halo are stacked
for all available outputs.}
\label{NDij}
\end{center}
\end{figure}

\begin{figure}
\begin{center}
\includegraphics[width=8cm]{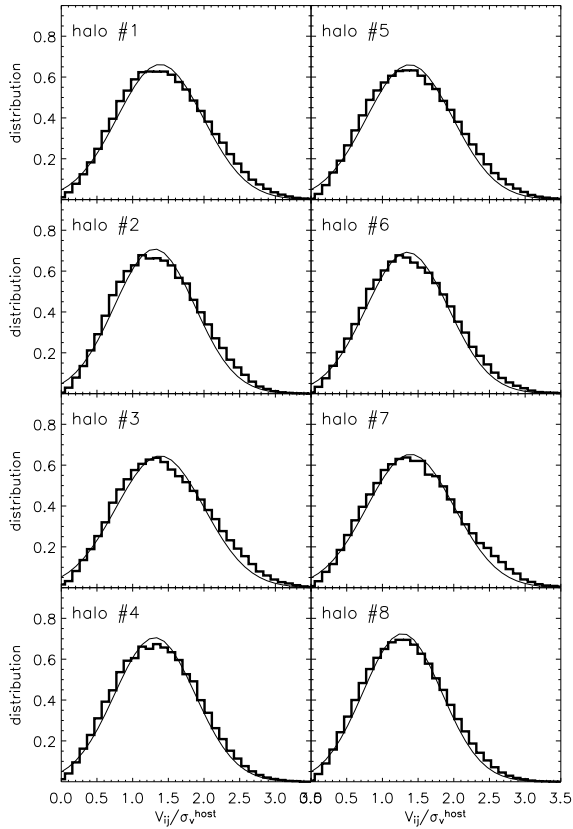}
\caption{Distribution of the relative velocity of two satellites
normalized by the host's velocity dispersion; for each host halo we 
stack the data from all available outputs.}
\label{NVij}
\end{center}
\end{figure}

\subsection{Distributions of Relative Encounters}

We have calculated the distribution of satellite-satellite distances 
$D_{i,j}$ as well as the relative speed of satellite pairs $V_{i,j,}$
for all available outputs in-between formation redshift $z_{\rm form}$
of the host and $z=0$, and show the resulting distributions of in
\Fig{NDij} and \Fig{NVij} respectively.

In \Fig{NDij} we have plotted the relative separation $D_{i,j}$
normalised by the sum of the two virial radii of the respective
satellites, i.e. $R_i$ and $R_j$; a value of $(R_i+R_j)/D_{i,j} > 1$ 
corresponds to a distance of the satellite$_i$-satellite$_j$ pair for 
which the ``virial spheres" of the satellites $i$ and $j$ are overlapping.
We note that the distributions can be fitted by a log-normal distribution

\begin{equation}
 n(x) = \displaystyle \frac{1}{x \sqrt{2\pi \sigma_0^{2}}} \exp \left( {-\frac{\ln^2 (x/x_0)}{2 \sigma_0^2}} \right)
\end{equation}

\noindent
where $x=(R_{i} + R_{j})/D_{i,j}$. \Fig{NDij} is accompanied by
\Table{BestFitNDij} where we summarize the best-fit parameters. 
Despite the age-IIM relation found in the previous
Section~\ref{IIMmeasure} we do not observe any trend for relative
distances to increase (or decrease) with halo age. \citet{TDS98}
performed a similar analysis, but their respective distance
distribution peaks for values corresponding to
distances smaller than the sum of the two individual radii indicating
they had ``at least one penetrating encounter" (cf. Fig.7 in their
paper noting that they are plotting the inverse of our distance
measure). However, we note that the definitition for a satellite's virial radius
used by \citet{TDS98} differs to ours; they define the virial radius to be the
satellite's radius at the moment it "merges" with the host halo, whereas we 
calculate the satellite's radius for each snapshot we have along its orbit 
within the host halo. This naturally leads to smaller radii as most of 
the satellites loses mass as it orbits within the denser environment of the 
host (c.f. definition for satellite radius in \Sec{SecIIM}, footnote 2),
and as a result the distribution of relative distances peaks at larger 
separations.

\begin{table}
\caption{Best fit parameters for relative distance distribution.}
\label{BestFitNDij}
\begin{tabular}{cccc}\hline
Halo         & $x_0$ & $\sigma_0$ \\
 \hline \hline

\#       1 &      0.097 &       0.536 \\
\#       2 &      0.111 &       0.510 \\
\#       3 &      0.158 &       0.567 \\
\#       4 &      0.130 &       0.550 \\
\#       5 &      0.104 &       0.568 \\
\#       6 &      0.137 &       0.564 \\
\#       7 &      0.147 &       0.543 \\
\#       8 &      0.119 &       0.540 \\

\hline
\end{tabular}
\end{table}

\begin{table}
\caption{Best fit parameters for relative velocity distribution.}
\label{BestFitNVij}
\begin{tabular}{ccc}\hline
Halo         & $w_0$ & $\sigma_0$ \\
 \hline \hline

\#       1 &        1.390 &       0.563 \\
\#       2 &        1.319 &       0.544 \\
\#       3 &        1.385 &       0.572 \\
\#       4 &        1.323 &       0.548 \\
\#       5 &        1.386 &       0.564 \\
\#       6 &        1.358 &       0.548 \\
\#       7 &        1.396 &       0.562 \\
\#       8 &        1.292 &       0.534 \\
\hline
\end{tabular}
\end{table}

Relative velocities between satellites can also enhance the impact of 
interactions on mass loss -- the slower the encounter between a pair of
satellites, the longer the timescale over which damage can be done.
In \Fig{NVij} we show the distribution of relative velocities for pairs
of satellites, normalised by the velocity dispersion of the host
halo. As before, we stack data for all available outputs for each
system, but we now fit the distributions with a Gaussian;

\begin{equation}
{ n(w) = \frac{1}{\sqrt{2 \pi \sigma^{2}}}  \exp \left(-\frac{(w-w_{0})^2}{(2 \sigma^2)} \right).}
\end{equation}

\noindent
Here $w = V_{\rm rel}/\sigma_{v}^{\rm host}$ is the relative velocity
of two satellites in terms of the velocity dispersion of the host
halo. This figure suggests that there is no correlation of peak value with
age, in good agreement with the best-fit parameters presented in
\Table{BestFitNVij}.

In summary, our analysis indicates that slow and/or close penetrating
encounters between pairs of satellite galaxies are relatively rare events.
We have checked to ensure that our decision to stack all available
outputs does not bias our result by masking a potentially interesting
signal; however, we can confirm that the results are unaffected whether
we construct the distribution from data obtained at a single redshift
(e.g. final redshift $z=0$ or formation redshift $z_{\rm form}$).

\begin{figure}
\begin{center}
\includegraphics[width=8cm]{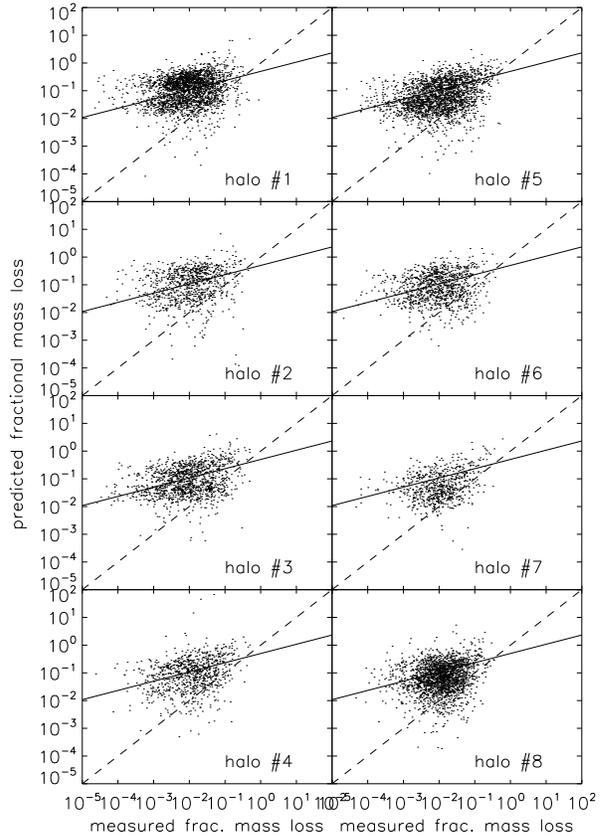}
\caption{Measured fractional mass loss in between two consecutive outputs
         versus the predicted mass loss as given by \Eq{Mloss}.
         The solid line represents a $x^{1/3}$ power law
         whereas the dashed line indicates a simple 1:1
         relation $x^{1.0}$.}
\label{MlossF}
\end{center}
\end{figure}

\subsection{Mass Loss induced by Satellite-Satellite Interactions}

We have defined a physically motivated quantitative measure of
interactions between satellite galaxies in the form of the IIM
(equation~\ref{IIM}). Our detailed investigation of IIM values in 
the previous sections provided great insight into the relevance of 
interactions in general; we were able to demonstrate that satellites
encounter other satellites with the same relative velocities and separations
at all times during the formation of a cluster, but the interaction measure
IIM is higher in younger systems. However, it is difficult to conceive 
of a means to reconstruct IIM values for satellite galaxies from observational
 data. Whereas IIM values can readily be evaluated in \nbody\ simulations 
providing a gauge for the presence and importance of interactions, 
respectively, we prefer to construct a new measure for quantifying the
impact of interactions more applicable to observational data sets. The mass 
loss suffered by a satellite as a result of the interactions would seem a
promising approach; it can be probed observationally, such as in
the field of ``galactic archaeology" where tidally stripped (stellar) 
streams have proven to be a powerful tool \citep[e.g.][]{HWdZ99}. However, 
understanding the evolution of satellite galaxies is complicated because
changes are driven not only by the tidal field of the host 
\citep[as shown by][]{KGKG05} but also by more subtle processes such as 
the time evolution of the underlying host potential. Explicitly accounting
for such time dependency gives better agreement with self-consistent
modeling of satellites in the integrals-of-motion space, but there still
remains a certain amount of disagreement between the observed and
measured mass losses; for example, \citet{KGKG05} speculated that
this can be attributed to either the shape of the host and/or
interactions with companion satellites. Using the ideas and
prescriptions developed in \Sec{IIMmeasure}, we now extend our analysis
to place constraints on the mass loss that can be induced by 
satellite-satellite interactions.\\

A satellite $i$ suffers a mass loss of 

\begin{equation}
 \Delta M^i = M^i(t_1) - M^i(t_2)
\end{equation}

\noindent
between two consecutive outputs $t_1$ and $t_2$. We wish to relate 
a fraction of this mass loss to interactions between satellites and so
we write $\Delta M^{i}$ as the sum of the mass loss induced by
interactions with other satellites, $\Delta M_{\rm sat}^i$, and with the
host halo, $\Delta M_{\rm host}^i$;

\begin{equation} \label{DeltaM}
\Delta M^i = \Delta M_{\rm sat}^i + \Delta M_{\rm host}^i \ .
\end{equation}

In order to break the degeneracy between $\Delta M_{\rm sat}^i$ and 
$\Delta M_{\rm host}^i$, we assume that the satellites are point masses 
$M$ with velocity $\vec{v}$, and hence write their momentum change 
$\Delta \vec{p}$ as follows:

\begin{equation}
 \Delta \vec{p} = \vec{v} \Delta M + M \Delta \vec{v} = \vec{F} \Delta t \ ,
\end{equation}

\noindent
which can be re-arranged to give

\begin{equation}
 \Delta M = (\vec{F}\cdot\vec{v}\Delta t - M\Delta\vec{v}\cdot\vec{v})/v^2 \ ,
\end{equation}

\noindent
where we (numerically) confirmed that on average \mbox{$\langle
 \vec{F}\cdot\vec{v}\Delta t \rangle \approx 7 \langle M\Delta\vec{v}\cdot\vec{v} \rangle$} and can therefore simplify the equation for mass loss to read

\begin{equation} \label{Mloss}
  \Delta M^{i} \propto \left(\frac{\vec{F}(t_{m}) \cdot \vec{v}(t_{m})}{v^{2}(t_{m})}  \Delta t \right)^{\alpha} \ ,
\end{equation}
 
\noindent
where $t_{m}$ is the midpoint between two outputs calculated using
\Eq{ForceMean} and $\alpha$ is a ``tuning factor'' accounting for the 
approximate nature of our approach.  From \Fig{ForceMean} we conclude that 
$\alpha \sim 1/3$ is the most appropriate value 
(represented by the solid line) and this is the value we adopt in the 
following analysis.

We now use \Eq{DeltaM} and the scaling relation \Eq{Mloss} to compute the mass
loss suffered as a result of satellite interactions and the influence of the 
host, respectively:

\begin{equation} \label{MLs}
 \begin{array}{lll} \Delta M_{\rm sat}^i & = & \displaystyle
 \frac{\Delta M^i}{1 + \left(\frac{ \vec{F}_{\rm host}^i(t_{m}) \cdot \vec{v}(t_{m}) }{ \vec{F}_{\rm sat}^i(t_{m}  \cdot \vec{v}(t_{m})) }\right)^\alpha } \\
\\
  \Delta M_{\rm host}^i & = & \displaystyle
 \frac{\Delta M^i}{1 + \left(\frac{ \vec{F}_{\rm sat}^i(t_{m})  \cdot \vec{v}(t_{m})}{ \vec{F}_{\rm host}^i(t_{m})  \cdot \vec{v}(t_{m}) }\right)^\alpha  }\\

 \end{array}
\end{equation}

\noindent
where we have further assumed that \Eq{Mloss} holds for both the force due
to satellite-satellite interactions and the force induced by the host
halo.

Over the years, a number of sophisticated prescriptions for modeling
tidally driven mass loss have been developed, largely within the context
of the evolution of globular clusters in external tidal fields \citep[e.g.][and references therein]{Spitzer,Gnedin} but also for understanding the 
disruption of satellite galaxies in cosmological dark matter halos
\citep[e.g.][]{Penarrubia,Hayashi,Taylor}. Although \Eq{MLs} represents a 
first order approximation for the mass loss, we will demonstrate that our 
formulae lead to qualitatively correct results and predictions with the right 
order of magnitude; a more thorough study and the development of a
full theoretical model for mass loss in cosmological dark matter halos
will be dealt with in a companion paper. In the present study we
concentrate on quantifying the importance of interactions for mass loss
and their importance for analytical modeling in galactic archaeology.

\begin{figure}
\begin{center}
\includegraphics[width=8cm]{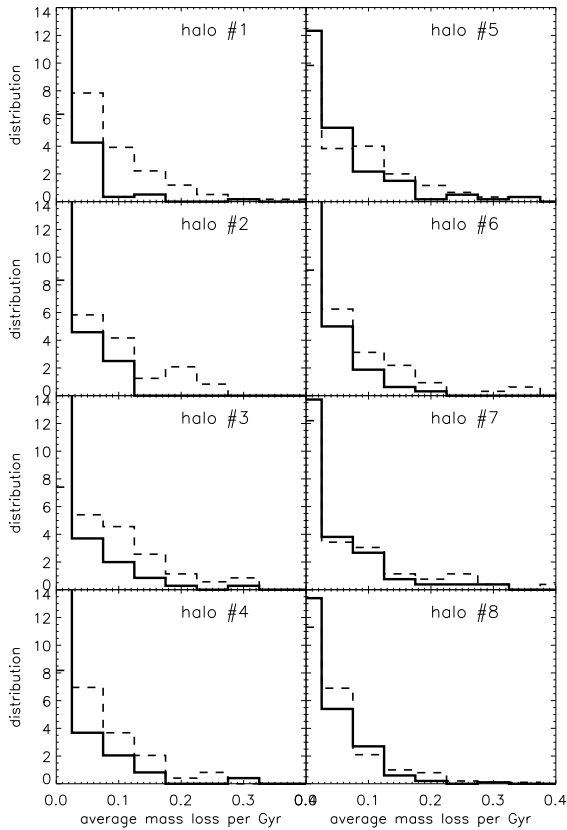}
\caption{Frequency distribution of the average (fractional) mass loss per Gyr.
        The thick solid line shows mass loss due to satellite interactions 
        and the dashed line due to the influence of the host alone.}
\label{NMLs}
\end{center}
\end{figure}

In the following analysis we use the average fractional mass loss per
Gyr for a given satellite $i$

\begin{equation}
\langle \frac{dM}{M dt} \rangle^i = \frac{1}{N_t^i}
  \sum_{t=t_i}^{t_{\rm now}} 
               \frac{M_t^i - M_{t-1}^i}{M_t^i} \frac{1}{\Delta t} 
\end{equation}

\noindent
where $N_t^i$ is the number of outputs available for that particular
satellite between the time it enters the host and the present; the time
interval $\Delta t$ is calculated for two consecutive outputs. We
are using \Eq{MLs} to split mass loss due to encounters and the
influence of the host. The resulting distributions for average mass
loss per Gyr are shown in \Fig{NMLs}. This figure demonstrated
that the mass loss induced by encounters between satellite galaxies can
be as important as the tidal stripping of mass by the host potential in
dynamically young systems. However, as the system becomes more relaxed,
the relevance of such interactions becomes progressively less
important and a significant fraction of the mass loss can be directly
ascribed to the tides induced by the host. \Table{MeanMassLoss}
accompanies \Fig{NMLs}; here we have calculated the mean of the average
mass loss per Gyr for all satellites in a given host halo:

\begin{equation}
\overline{\langle \frac{dM}{M dt} \rangle} =
\frac{1}{N_{\rm sat}} \sum_{i=1}^{N_{\rm sat}} 
\langle \frac{dM}{M dt} \rangle^{i}
\end{equation}

\begin{table}
\caption{The mean average fractional mass loss per Gyr when averaging
         over all times and all satellites in a halo. The last column
         measures the contribution from satellite-satellite interactions,
         i.e. $f_{\rm sat}=\overline{\langle \frac{dM}{M dt} \rangle_{\rm sat}}/\overline{\langle \frac{dM}{M dt} \rangle_{\rm total}}$}
\label{MeanMassLoss}
\begin{tabular}{ccccc}\hline
halo        & $\overline{\langle \frac{dM}{M dt} \rangle_{\rm total}}$ 
               & $\overline{\langle \frac{dM}{M dt} \rangle_{\rm sat}}$
               & $\overline{\langle \frac{dM}{M dt} \rangle_{\rm host}}$ 
               & $f_{\rm sat}$\\
 \hline \hline

\#       1 &      0.132 &     0.030 &     0.102 &       0.23 \\
\#       2 &      0.123 &     0.033 &     0.090 &       0.27 \\
\#       3 &      0.143 &     0.039 &     0.104 &       0.27 \\
\#       4 &      0.126 &     0.038 &     0.088 &       0.30 \\
\#       5 &      0.147 &     0.050 &     0.097 &       0.34 \\
\#       6 &      0.137 &     0.041 &     0.096 &       0.30 \\
\#       7 &      0.141 &     0.054 &     0.087 &       0.38 \\
\#       8 &      0.123 &     0.055 &     0.068 &       0.45 \\

\hline
\end{tabular}
\end{table}

From Table~\ref{MeanMassLoss} we infer that the mass loss induced by
satellite-satellite encounters can amount to as much as 45\% of the
total mass loss experienced by a single satellite. Even though the
integral interaction measure IIM (as defined by \Eq{IIM}) and its
distribution in \Fig{NIIM} indicated a rather low importance of such
interactions the conversion to mass loss reveals a more pronounced
influence due to the observed power law scaling $\Delta M \propto (F
\Delta t)^{\alpha}$ with $\alpha \sim 1/3$. However, the results are
robust to changes in the power-law index, e.g. changing the exponent 
from $1/3$ to unity gives us the range from 15\% mass loss due to 
interactions for the oldest host (halo~\#1) up to 40\% for the youngest
system (halo~\#8).



\subsection{A Test Scenario: Host Halo~\#8} \label{TestCase}

Two questions remain unanswered: 
\begin{enumerate}
\item \label{1} Why do we observe a higher mass loss due to
interactions in younger systems?
\item \label{2} Can our results be verified?  
\end{enumerate}
The most natural approach to addressing these questions involves 
explicitly tracking the mass loss of an individual satellite as a
function of time and factoring out the influence of the other satellites.
To do this, we have performed two additional simulation runs of halo~\#8,
 both starting at its formation redshift $z=0.3$. In the first, we have 
removed all halos bar the progenitor of the $z=0$ host halo and one particular
 satellite that happened to have a rather high interaction value of
 $\IIM = 8.3$; our analysis indicates that about 40\% of its average mass 
loss per Gyr was induced by interactions with other satellites. In the second
 run, we removed only those subhaloes that have not merged with the host's
progenitor at redshift $z=0$ except our ``test satellite". We refer to these 
runs as ``fully cleaned" (the former, including \textit{only} the host and our
 test satellite) and ``cleaned" (the latter, also including the massive
 subhaloes merging with the host) respectively.

A visual impression of the initial setup of the cleaned run is given
in (the upper left panel of) \Fig{sat55orbit} which nicely demonstrate the 
``smoothness" of the cleaned simulation. The satellite in question is marked
 by a blue circle.

\begin{figure}
\begin{center}
\includegraphics[width=8cm]{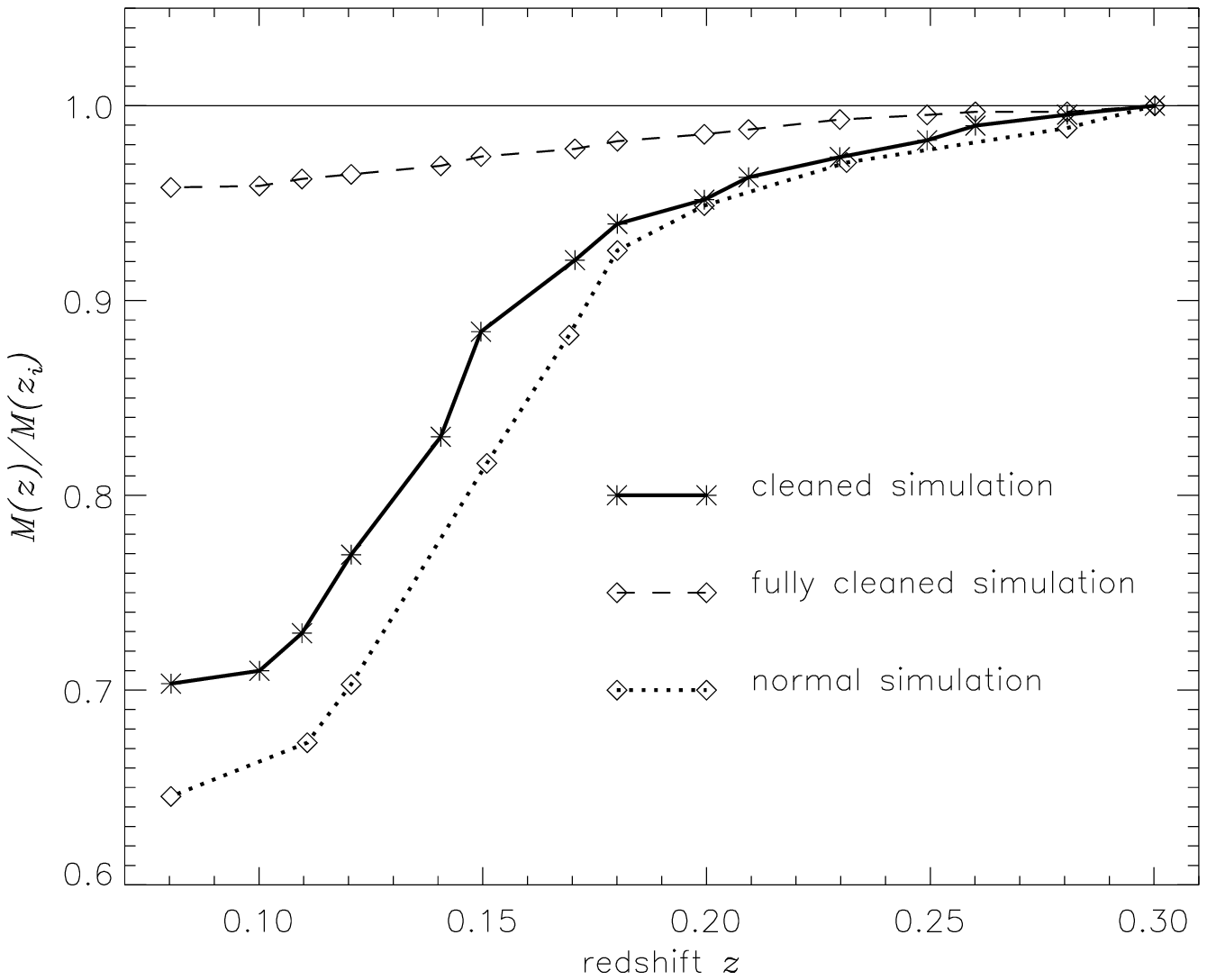}
\caption{Tracking the mass of the subhaloes indicated by the
         red circle in \Fig{sat55orbit} from initial redshift 
         $z=0.31$ to $z=0.08$ through the actual (``normal")
         simulation as well as out two test cases described
         in \Sec{TestCase}.}
\label{sat55MassHistory}
\end{center}
\end{figure}

For each of the three runs we closely follow the mass loss history of
this satellite and the resulting curve (normalized to the initial
mass) is presented in \Fig{sat55MassHistory}. Note that we identify the 
set of particles that are bound to the satellite at the initial time and
 we explicitly track these particles through subsequent snapshots, checking 
what fraction are bound to the satellite at later times. This avoids any
difficulties that may arise from attempting to identify the bound mass of 
the satellite as it is identified in subhalo catalogues constructed for 
consecutive snapshots. For each available output we find the new satellite 
centre by using the centre-of-mass of the innermost
bound particles as a first guess for the central density peak. We then
iteratively remove all of the satellite's particles that are not
gravitationally bound. \Fig{sat55MassHistory} shows that the mass loss suffered
by the satellite is significantly reduced in both of the cleaned runs.  
However, we stress that removing all substructure not only affected the 
satellite under investigation but also the overall dynamics of the host, 
especially for the fully cleaned run: host halo~\#8 can be classified as a
violent (triple) merger in the original cosmological simulation and
removing all of its progenitors must clearly leave an imprint on
its (internal) dynamics. Nevertheless, we note that although the ``cleaned'' 
run retained most of its high mass substructures, we observed a trend for
the mass loss to decrease.

\begin{figure}
\begin{center}
\includegraphics[width=8cm]{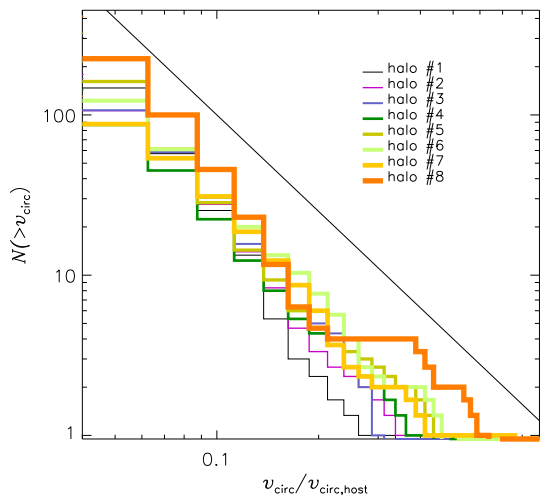}
\caption{Cumulative circular velocity distribution for all subhaloes in
         the eight host halos half way through the evolution
         from redshift $z=z_{\rm form}$ to $z=0$. The solid line
         represent the power law $x^{-2}$.}
\label{Nvcirc}
\end{center}
\end{figure}

\begin{figure*}
\begin{center}
\includegraphics[width=14cm]{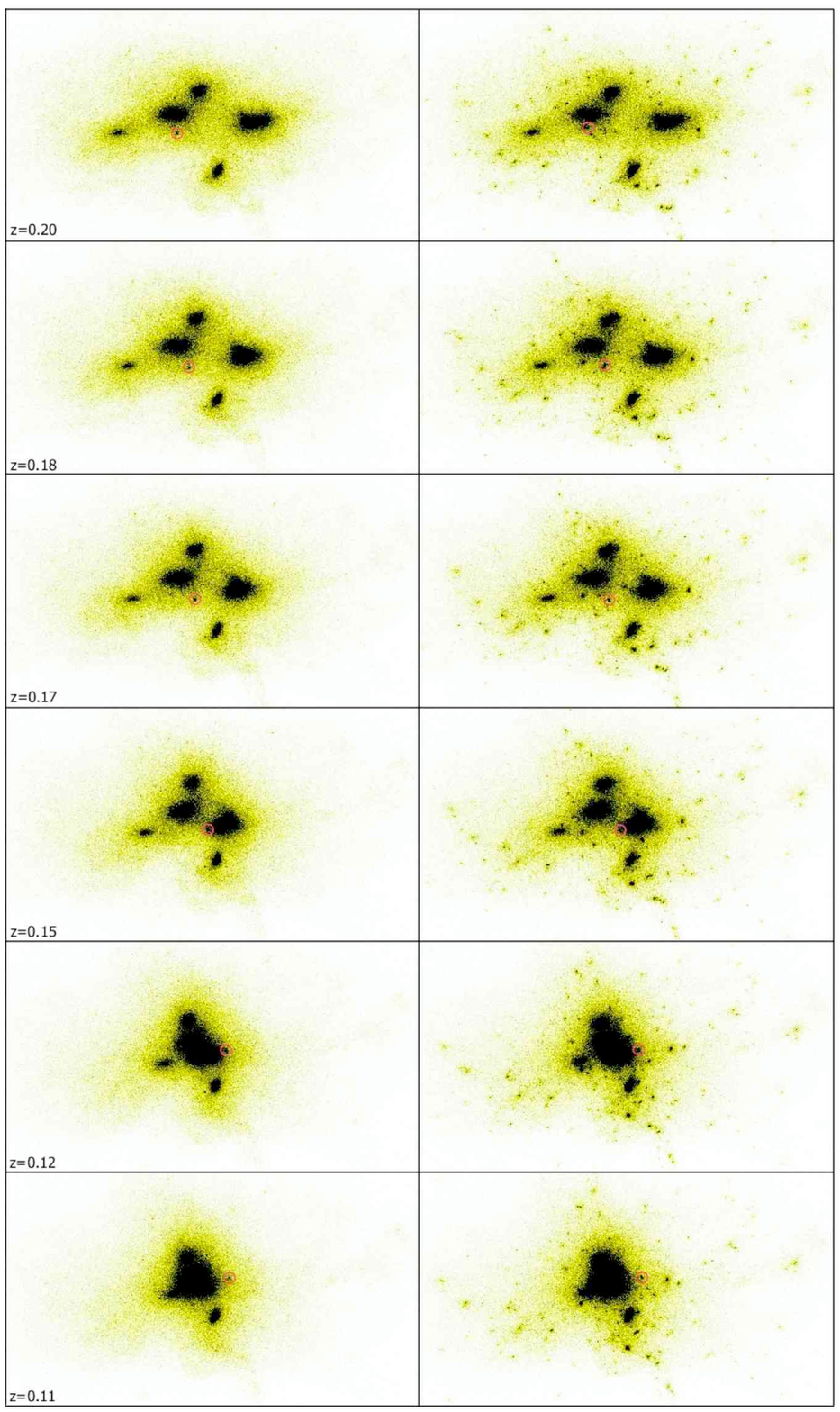}
\caption{Several snapshots of the actual simulation of
         halo~\# 8 (right panel) and the re-simulation
         cleaned of all (but one) subhalo not ending up in
         the host at $z=0$. The subhalo visible to the
         lower right of the host happens to be a foreground
         objects not interfering with the system under
         investigation.}
\label{sat55orbit}
\end{center}
\end{figure*}

This simple test has demonstrated that the principal driver for interaction
induced mass loss is the mass spectrum of the substructure halos.
We may strengthen this statement further by considering the cumulative 
circular velocity distribution presented in \Fig{Nvcirc}. Noting that
the maximum circular velocity is a reasonable measure of halo mass,
we conclude from this figure that a sizeable fraction of the mass in 
satellites in each of the eight host haloes is bound to high mass systems 
roughly half way through their evolution (i.e. $t=(t_{\rm form}+t_0)/2$). 

These massive subhaloes are likely to be responsible for the increased mass 
loss due to interactions, as already indicated by \Fig{sat55MassHistory}. We
can investigate this assertion by determining the fraction of the IIM measure
that is due to massive satellites. As already mentioned in \Sec{IIMmeasure}, 
we can reproduce the distributions of IIM values presented in \Fig{NIIM} 
by including only those satellite galaxies that are more massive than 1\% 
of the host's virial mass\footnote{Using an empirically derived scaling 
relation between satellite mass and maximum circular velocity, i.e. 
$v_{\rm circ} \propto M^{1/3}$, the mass limit of 1\% $M_{\rm vir, host}$ 
corresponds to a cut in circular velocity at around 20\% $v_{\rm circ, host}$
 (cf. \Fig{Nvcirc}).}. This indicates quite clearly that the IIM values are 
dominated by more massive systems and the contribution of low-mass satellites 
is negligible. 

\section{Conclusions} 
\label{Conclusions}

The hierarchical manner in which structure in our Universe forms -- 
from the bottom up, through mergers and accretions -- implies that 
interactions between galaxies (and consequently, between their dark 
matter haloes) are commonplace. These interactions have been invoked 
to explain, for example, galaxy transformation \citep[e.g.][]{MLK98}, 
the exchange of angular momentum \citep[e.g.][]{B87}, the triggering 
of star bursts \citep[e.g.][]{Mihos} and morphological change
\citep[][]{Steinmetz}. The impact of interactions between satellite
galaxies and their massive host on the structure of the satellites has been
studied in some detail \citep[e.g.][]{Hayashi}, but less well understood is the
role played by interactions between satellite galaxies; in other words, the 
impact of satellite-satellite interactions. This important topic has formed 
the basis of this paper.

We have defined an \emph{integral interaction measure} (IIM) that
allows us to quantitatively measure the importance of interactions
between satellite galaxies for their mass loss. Our definition allows
us to gauge the relative contributions of the host potential and other
satellites for the mass loss suffered by an individual satellite. We
have shown that the distribution of IIMs for a population of satellites
within a cluster mass dark matter halo can be characterised as
lognormal, and that the peak value (or mode) correlates with the age of
the host system -- typically the younger the host, the larger the peak
IIM. Moreover, we note that the relative width of the distribution is
broader in younger systems. We were able to confirm that the most
significant contribution to the interaction measure comes from massive
companion satellites which naturally explain the correlation with host age:
subhaloes in young clusters have larger masses relative to the host since they
have not been tidally disrupted yet which is validated by the observation that
our younger hosts have a higher mass fraction in satellites. However, the IIM
values are generally much less than unity, implying that the bulk of the mass
loss suffered by a satellite is driven by its interaction with the host 
potential. We have also shown that, in those cases where the IIM is large, 
it cannot be due to single encounters; rather, it is built up through a
 series of many encounters.

Our investigations have also extended the result of \citet{KGG04}
by demonstrating that not only are penetrating encounters between
satellite galaxies relatively rare events over the ``lifetime'' of a 
cluster\footnote{i.e. since its formation redshift.}, but that the
timescale of such encounters is short, i.e. the relative velocities are
typically of order the 1D velocity dispersion of the host. This result
may be of interest to those engaged in developing semi-analytic models
of galaxy formation because we might expect the severity of encounters
between satellites to be important for the efficiency of starbursts
arising from tidal interactions. 

Finally we have proposed a simple \emph{empirical} model for separating
the respective contributions of the host potential and interactions
with other satellites for the mass loss suffered by a satellite. Our
model suggests that mass loss driven by satellite interactions can be 
significant -- in the particular test case we considered, we have shown
that a given satellite can lose as much as $\sim 40\%$ of its initial mass.
This may appear surprising at first, but we have shown that the IIM is
a cumulative measure and so while damaging encounters are relatively
rare occurrences, a large number of ``weak'' interactions can affect
the structure of a satellite and drive the mass loss it
suffers. However, we stress that this empirical model should be taken as
simple guide providing an ``order-of-magnitude'' estimate of the mass
loss, and a more sophisticated model is required; this will be the
focus of future work.

\section{Acknowledgments} 
The simulations presented in this paper were carried out on the
Beowulf cluster at the Centre for Astrophysics~\& Supercomputing,
Swinburne University. AK acknowledges funding through the Emmy Noether Programme by the DFG (KN 722/1). CP thanks Virginia Kilborn and Sarah Brough
for helpful discussion. The financial support of the Australian Research 
Council is gratefully acknowledged.

\label{lastpage}

\end{document}